\documentclass[fleqn,usenatbib,useAMS]{mnras}

\usepackage{graphicx}	
\usepackage{amsmath}	
\usepackage{amssymb}	
\usepackage{multicol}        
\usepackage{bm}		
\usepackage{pdflscape}	

\usepackage[T1]{fontenc}
\usepackage{ae,aecompl}

\defcitealias{Shankar:2016}{Sh$+$16}

\title[Quashing a suspected selection bias]{Quashing a suspected selection bias in galaxy samples having dynamically-measured supermassive black holes}

\author[Sahu, Graham, \& Hon]
{
Nandini Sahu$^{1,2}$
\thanks{Contact e-mail: \href{mailto:nandinisahu2705@gmail.com}{nandinisahu2705@gmail.com}}
Alister W.\ Graham$^2$ 
Dexter S.-H. Hon$^2$
\\
$^1$ OzGrav-Swinburne, Centre for Astrophysics and Supercomputing, Swinburne
University of Technology, Hawthorn, VIC 3122, Australia\\
$^2$ Centre for Astrophysics and Supercomputing, Swinburne University of
Technology, Hawthorn, VIC 3122, Australia\\ 
}

\date{Accepted XXX. Received YYY; in original form ZZZ}
\pubyear{2022}

\begin{document}
\label{firstpage}
\pagerange{\pageref{firstpage}--\pageref{lastpage}}
\maketitle

\begin{abstract}

Local early-type galaxies with directly-measured black hole masses, $M_{\rm
  bh}$, have been reported to represent a biased sample relative to the
population at large.
%
%
Such galaxies with Spitzer Space Telescope imaging have been purported to
possess velocity dispersions, $\sigma$, at least $\sim$0.1 dex larger for a
given galaxy stellar mass, $M_{\rm *,gal}$, than is typically observed among
thousands of early-type galaxies imaged by the Sloan Digital Sky Survey.  This
apparent offset led Shankar et al.\ to reduce the normalisation of the
observed $M_{\rm bh} \propto \sigma^5$ relation by at least $\sim$0.5 dex to
give their ``intrinsic relations'', including $\sigma$-based modifications to
the $M_{\rm bh}$--$M_{\rm *,gal}$ relation.  These modifications were based on
the untested assumption that the stellar masses had been derived consistently
between the two samples.  Here, we provide the necessary check using galaxies
common to the Spitzer Survey of Stellar Structure in Galaxies (S$^4$G) and the
Sloan Digital Sky Survey (SDSS).  We find that the stellar masses of galaxies
with and without directly measured black holes had appeared offset from each
other due to the use of inconsistent stellar mass-to-light ratios,
$\Upsilon_*$, for the optical and infrared data.  We briefly discuss the
``intrinsic relations'' and why some of these will at times appear
to have had partial success when applied to data based on similarly inconsistent
values of $\Upsilon_*$.  Finally, we reiterate the importance of the $\upsilon$
(lower-case $\Upsilon$) term, which we previously introduced into the $M_{\rm
  bh}$--$M_{\rm *}$ relations to help avoid $\Upsilon_*$-related mismatches.

\end{abstract}



\section{Introduction}
\label{introduction}

The consensus regarding the presence of a supermassive black hole (SMBH) at
the centre of almost every galaxy (local/quiescent or distant/active) stems
from multiple pioneering studies, e.g., \citet{Baade:Minkowski:1954,
  Hoyle:Fowler:1963, Burbidge:Burbidge:Sandage:1963, Salpeter:1964,
  Zeldovic:1964, Schmidt:1965, Sandage:1965, Lynden:1969,
  Lynden:Rees1971}. \citet{Dressler:Richstone:1988} were the first to clearly
suggest a correlation between black hole mass ($M_{\rm bh}$) and the host
galaxy's bulge mass (using merely two galaxies).  With an ever-increasing
number of direct dynamical\footnote{Direct SMBH mass measurement methods include
  stellar and gas dynamical modelling, megamaser kinematics, proper motion
  (for Sgr A$^*$) and direct imaging (for M87).} SMBH mass measurements ---
tabulated in, for example, \citet{Ferrarese:Ford:2005},
\citet{Kormendy:Ho:2013}, and \citet{Sahu:2019:II} ---, the study of BH--galaxy
correlations has been actively pursued \citep{Dressler:1989, Yee:1992,
  Kormendy:Richstone:1995, Magorrian:1998, Laor:1998, Wandel:1999,
  Salucci:2000, Laor:2001, 2002ApJ...578...90F, Haring:Rix:2004, Ferrarese:Ford:2005,
  Graham:2007:Mbh, Gultekin:Richstone:2009, Sani:2011, 
  Beifiori:Courteau:2012, Graham:2012, Kormendy:Ho:2013, McConnell:Ma:2013,
  Graham:Scott:2013, Graham:Scott:2015, Lasker:2014, Saglia:2016,
  Savorgnan:2016:Slopes, Davis:2018:a, Sahu:2019:II}.  Although an evolving
field, it is now quite clear that spheroids built from mergers which have
folded in the progenitor galaxies' disc mass form an offset sequence in the
$M_{\rm bh}$--$M_{\rm *,bulge}$ diagram, possessing lower $M_{\rm bh}/M_{\rm
  *,bulge}$ ratios \citep{Sahu:2019:I, Graham:Sahu:2022}.\footnote{A review of
  how the perception of the existence of SMBHs has evolved can be found in
  \citet{Shields:1999} and \citet{Genzel:2021}, while developments in the
  observed BH-galaxy correlations (until 2015) are additionally reviewed in
  \citet{Graham:2016:Review}.}

\citet[][their Figure~1]{2007ApJ...660..267B}, see also 
\citet{2002MNRAS.335..965Y}, claimed
that the sample of galaxies with directly measured black hole masses had
overly-large velocity dispersions at a given bulge magnitude relative to a
larger SDSS sample of ETGs.  That claim, explored in the Appendix of 
\citet{2007MNRAS.379..711G}, 
was based on applying a single, two-parameter 
de Vaucouleurs' $R^{1/4}$ model fit to the ETGs in the SDSS sample, potentially
overestimating the brightness of the actual bulge component.   A similar claim
that the sample of galaxies with directly measured black hole masses is
biased, such that they have overly-large velocity dispersions at a given
galaxy stellar mass, relative to the SDSS sample, was made by 
\citet[][ \citetalias{Shankar:2016} hereafter]{Shankar:2016} 
but this time armed with improved S\'ersic-bulge plus
exponential-disc decompositions for the SDSS sample. The benefit was
that the five-parameter fits to the SDSS data provided a better estimate of
the total galaxy magnitude. 
This claim was based upon an offset seen between
samples with and without SMBH mass measured directly/dynamically in the (stellar
velocity dispersion: $\sigma$)--(galaxy stellar mass: $M_{\rm *, gal}$)
diagram.  They suggested it occurred due to a resolution-limited selection
effect in the dynamically-measured black hole sample.

We investigate this apparent offset in the $\sigma$--$M_{\rm *, gal}$ diagram,
which may have significant ramifications for SMBH scaling relations and their
application.  Some important applications include calibration of the
virial factor for the reverberation mapping of (distant) active galaxies,
insights into SMBH--galaxy coevolution theories, tests for galaxy
simulations, derivation of the SMBH mass function, estimation of SMBH merger
timescales, SMBH merger rate, and the expected amplitude of the long-wavelength
gravitational wave (GW) signals\footnote{See \citet{Sahu:2022} for details on
  how the latest SMBH scaling relations can obtain SMBH merger
  timescales, merger rate estimates and significantly improve the modelling
  of the expected long-wavelength GWs.} looked for by pulsar timing arrays and
future space interferometers.


\begin{figure}
\begin{center}
\includegraphics[clip=true,trim= 10mm 08mm 20mm 20mm,width=  \columnwidth]{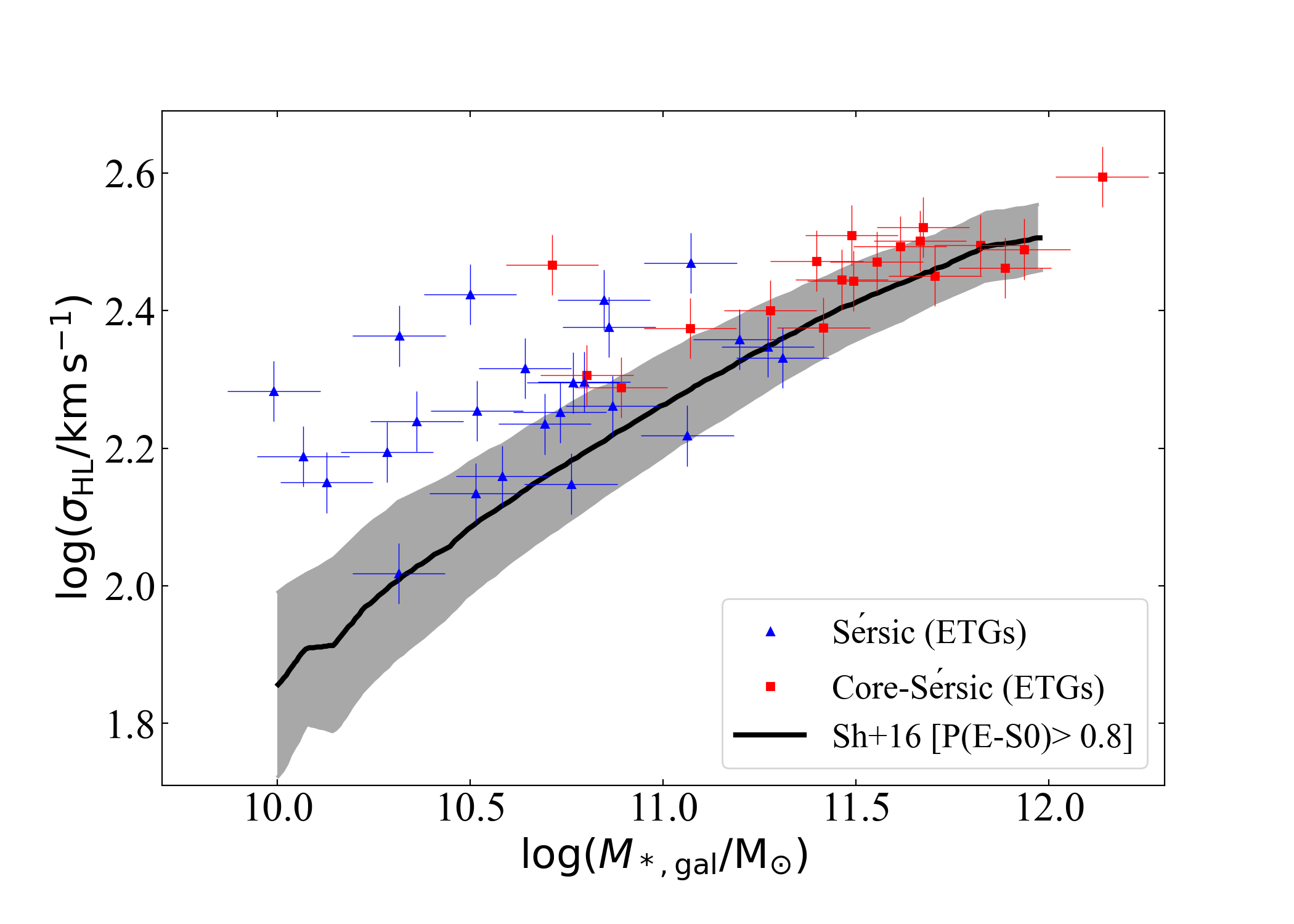} 
\caption{An updated version of the $\sigma$--$M_{\rm *,gal}$ diagram from
  \citetalias{Shankar:2016} (top right panel of their Figure~1) comparing ETGs
  with and without a dynamical SMBH mass measurement.  The black curve, 
\citepalias[taken from][their Figure~1]{Shankar:2016}, is defined by a sample of
  ETGs from SDSS-DR7 that are currently without a dynamical SMBH mass
  measurement.  The grey band, also taken from \citetalias{Shankar:2016},
  outlines the 68 per cent scatter region.  ETGs with a dynamical SMBH mass
  measurement (marked by red squares and blue triangles) are from
  \citet{Savorgnan:2016:Slopes}.  This figure shows 44 ETGs from the 
  \citet{Savorgnan:2016:Slopes} catalogue compared to the 37 ETGs used in
  \citetalias{Shankar:2016} because stellar velocity dispersions
  for the remaining 7 were not available at that time.  The two samples are
  described in Section \ref{Data}.  }
\label{Sigma_Mgal_sh}
\end{center}
\end{figure}

\citetalias[][their Figure~1]{Shankar:2016} compared local early-type
galaxies (ETGs) which have dynamical SMBH mass measurement, specifically, the BH
samples from four different studies \citep{Beifiori:Courteau:2012,
  McConnell:Ma:2013, Lasker:2014, Savorgnan:2016:Slopes} with a sample of ETGs
from Data Release 7 (DR7) of the Sloan Digital Sky Survey \citep[hereafter
  SDSS-DR7,][]{Abazajian:2009}.  They showed an offset in all four cases, the
latest of which is presented in Figure~\ref{Sigma_Mgal_sh}, where the
$\sigma$--$M_{\rm *,gal}$ curve formed by the SDSS-DR7 ETG sample is compared
with 44 ETGs (with direct SMBH mass measurement) from
\citet{Savorgnan:2016:Slopes}.  Here, the galaxy morphology class ``ETG"
refers to elliptical (E), ellicular (ES)\footnote{Ellicular (ES-type) refers
  to ETGs with intermediate-scale rotating stellar disks within their dominant
  spheroidal component. Such galaxies have been known since
  \citet{Liller:1966}.  The term ``ellicular" combines elliptical and
  lenticular, and ES-type galaxies are placed between the E- and S0-types in
  the morphology classification grid presented by \citet{Graham:Grid:2019}.},
and lenticular (S0) galaxies, whereas the late-type galaxy (LTG) refers to
spiral (S) and irregular (Irr) galaxies.

The velocity dispersions for the \citet{Savorgnan:2016:Slopes} SMBH sample
were taken from the \textsc{HyperLeda}
database\footnote{\url{http://leda.univ-lyon1.fr/}} \citep{Paturel:2003,
  Makarov2014}, which are homogenised at an aperture (radius) size of R=0.595
kpc.  Often referred to as the central velocity dispersion, it is denoted here
as $\sigma_{\rm HL}$.  The black curve shown in Figure~\ref{Sigma_Mgal_sh} is
directly taken from \citetalias{Shankar:2016} (top right panel of their
Figure~1).  It represents the mean $\sigma_{\rm HL}$--$M_{\rm *,gal}$ relation
defined by the ETGs from the SDSS-DR7 sample, and the grey shaded regions mark
the 68 per cent scatter of the distribution around the mean (black) curve.  The
SMBH sample shown in Figure~\ref{Sigma_Mgal_sh} has seven extra ETGs from the
\citet{Savorgnan:2016:Slopes} sample for whom $\sigma_{\rm HL}$ is now
available from the \textsc{HyperLeda} database.
 
\citetalias{Shankar:2016} selected the ETGs from the SDSS-DR7 sample using a
Bayesian automated morphological classifier \citep{Huertas-Company:2011},
where galaxies with a probability of greater than 0.8 for being E- or
S0-types, denoted by $p(\rm E-S0) > 0.8$, were placed in the ETG bin.  The
classifier algorithm in \citet{Huertas-Company:2011} assigned a probability to
each galaxy for belonging to four different morphological types: E
(elliptical); S0 (lenticular); Sab (early LTG); and Scd (late LTG).
\citet[][in their Section 6]{Sahu:2019:II} argued that $p(\rm E-S0) > 0.8$
implied that at least 10 per cent of the SDSS-DR7 ETGs selected by
\citetalias{Shankar:2016} using the above criteria were misclassified and may
be LTGs (Sab or Scd).  \citet{Sahu:2019:II} noted that the apparent offset of
the SMBH sample from the SDSS-DR7 ETG $\sigma_{\rm HL}$--$M_{\rm *,gal}$ curve
seems higher at the low-mass end ($10 \lesssim \log(M_{\rm *,gal}/{\rm
  M}_{\odot}) \lesssim 10.5$) than at the high masses due to at least 10 per cent
contamination by LTGs in the supposed ETG sample of \citetalias{Shankar:2016}.
This is because the LTGs (spiral) define a different 
$\sigma_{\rm HL}$--$M_{\rm *,gal}$ relation than ETGs and reside below ETGs in the
$\sigma_{\rm HL}$--$M_{\rm *,gal}$ diagram \citep[see Figure~15
  in][]{Sahu:2019:II}.  This may be a reason behind a fraction of the offset 
between the ETG SMBH sample and the SDSS-DR7 ETG sample.  However,
\citet{Sahu:2019:II} did not thoroughly investigate this offset which is now
done here.\footnote{In 2019, we were aware of the results presented herein; 
  however, pandemic-related delays have meant that we are only reporting them
  now.}

Importantly, from the SDSS-DR7 spectroscopic galaxy sample without a
direct-dynamical SMBH mass measurement, \citetalias{Shankar:2016} used the
ETGs within the redshift range of $0.05 <z< 0.2$.  The distance of their
sample meant that they had no galaxies in common with the sample containing
direct dynamical SMBH mass measurements. This prohibited them from checking if
the stellar masses were being derived consistently between the two samples.
\citetalias{Shankar:2016} interpreted
the offset seen in the $\sigma_{\rm HL}$--$M_{\rm *,gal}$ diagram 
in terms of the SMBH sample having a higher
velocity dispersion compared to the majority of galaxies with a similar galaxy
mass but which do not have their central SMBH mass directly dynamically
measured. However, another potential reason behind this apparent offset is a
discrepancy in galaxy stellar mass.

For our investigation of a mass discrepancy between the SMBH sample and the
larger SDSS sample, we use i) a 
sample of local galaxies (without SMBH mass measurement but) with SDSS
imaging, analysed with the same techniques as the SMBH sample, and ii) data
from the Spitzer Survey of Stellar Structure in Galaxies
\citep[$S^4G$,][]{Sheth:2010}, both of which we can use as a reference sample for mass
comparison.  Essentially, we have now performed the same (consistent) galaxy
image analysis/decomposition on a sample of $\sim$100 local galaxies from the
SDSS and $\sim$100 local galaxies (having directly measured SMBH masses) 
imaged with the Spitzer Space Telescope. Cross comparison with each other, with the
larger SDDS sample, and the $S^4G$ sample, enables us to establish if the
stellar masses have been determined consistently.

In Section~\ref{Data}, we describe the data used for our test.  Section
\ref{analysis} compares the SDSS-DR7 $\sigma_{\rm
  HL}$--$M_{\rm *,gal}$ curve with the $\sigma_{\rm HL}$--$M_{\rm *,gal}$
relations defined by the updated SMBH sample before we resolve the reason
behind the current apparent offset in the $\sigma_{\rm HL}$--$M_{\rm *,gal}$
diagram.  Section \ref{Implication for BH} discusses the implication of our
investigation for the alleged bias in the SMBH scaling relations defined by
the dynamically-measured SMBH sample, and Section \ref{conclusions} summarises
the findings and the broader implications of this investigation.

\section{Data} 
\label{Data}

We calculate the stellar masses of our local SDSS sample of $\sim$100 galaxies
following the same stellar mass-to-light ratio prescription used in
\citetalias{Shankar:2016} for the larger SDSS-DR7 sample.  The stellar
mass-to-light ratio used for the SMBH sample with $3.6\,\mu$m imaging is
similar to that used for the $S^4G$ sample. By comparing the stellar masses of
the two samples with the values independently obtained from the $S^4G$ survey
for the same galaxies, we can establish if there is an offset in the mass
derivation, rather than a biased sample, between the sample with direct SMBH
mass measurements and the population at large.  A description of these three
samples, along with more details on $\sigma_{\rm HL}$ and $M_{\rm *,gal}$ for
the SDSS-DR7 sample, is provided here.

\subsection{SMBH sample}
\label{BHS}

The largest-to-date sample of galaxies with dynamically (directly) measured
black hole mass currently numbers $\sim$150 \citep[145 galaxies are listed
  in][]{Sahu:2019:II}.  For 127 of these, we have measured the host
galaxy properties from careful image analysis, advanced isophotal modelling 
\citep[in terms of capturing structures and 
  irregularities,][]{Ciambur:2015:Ellipse}, and physically-motivated multicomponent
decomposition of the galaxy light.  The complete image analysis of the 127
galaxies, primarily (81 per cent) using 3.6 $\mu$m Spitzer Space Telescope images,
was collectively done by \citet{Savorgnan:Graham:2016:I},
\citet{Davis:2018:a}, and \citet{Sahu:2019:I}, and readers are directed to
these parent studies for a detailed description of the galaxy modelling and
decomposition procedures, and references to the sources which measured
the SMBH mass.  There are 73 ETGs and 28 LTGs with Spitzer imaging. 

Here, we refer to this reduced Black Hole Sample in ETGs with Spitzer imaging
as BHS, and we use the galaxy magnitudes and morphologies in
\citet{Sahu:2019:I} and \citet{Savorgnan:Graham:2016:I}.  The $M_{\rm *,gal}$
for the BHS were calculated using a constant stellar mass-to-light ratio of
0.6 \citep[following][]{Meidt:2014} based upon the Chabrier IMF, luminosity
distances \citep[provided for the whole BHS in][Table~1]{Sahu:2019:II}, and
the galaxy magnitudes measured in above studies.


The central velocity dispersion for the BHS is primarily taken from the
\textsc{HyperLeda} database, already provided in
\citet[][Table~1]{Sahu:2019:II}, where other sources for a few galaxies are
also specified there.

\subsection{SDSS-DR7 sample}
\label{SDSS_DR7}

This is the sample used by \citetalias{Shankar:2016}. 
The galaxy luminosities and effective half-light radii ($R_{\rm e}$) of the
SDSS-DR7 sample come from the two-component (bulge plus disk, \textsc{SerExp})
decompositions of the galaxy light obtained from SDSS $r$-band
images analysed in \citet{Meert:Vikram:2015}.  Following \citet[][their
  Equation 6]{Bernardi:2010}, \citetalias{Shankar:2016} converted the K- and
evolution-corrected galaxy luminosity to galaxy stellar mass using a ($g-r$
colour)-dependent stellar mass-to-light ratio from \citet{Bell:2003} based upon
the \citet{Chabrier:2003} Initial Mass Function (IMF) of stars.
\citetalias{Shankar:2016} denoted galaxy stellar mass by $M_{\rm STAR}$;
however, we will denote galaxy stellar mass by $M_{\rm *,gal}$ for consistency
with notation in our previous works and to avoid ambiguity between bulge and
galaxy stellar mass.

SDSS provides stellar velocity dispersions measured at an aperture size of
$R_{\rm e}/8$ for a galaxy.  In order to obtain velocity dispersions
consistent with the HyperLeda values homogenised at an aperture size of 0.595
kpc, which was/are also used for the SMBH sample, \citetalias{Shankar:2016}
applied an empirical mean aperture correction\footnote{$\sigma_R/\sigma_e =
  (R/R_e)^{-0.066}$.} provided in \citet{Cappellari:2006} for converting the
SDSS $\sigma_{R_{\rm e}/8}$ values to $\sigma_{\rm HL}$.

\subsection{SDSS-$i$ sample}
\label{SDSS_i}

This sample of about 100 local galaxies, which we refer to as the ``SDSS-$i$
sample" throughout this paper, uses galaxy magnitudes measured by
\citet{Hon:2022} from $i$-band images taken from the SDSS Data Release 8
(DR8).  Similar to the SDSS-DR7 sample, they do not yet have a direct SMBH
measurement. \citet{Hon:2022} used the same image analysis techniques as was
done for the sample with black hole measurements (Section \ref{BHS}).
\citet{Hon:2022} 
also performed two-component decomposition for their sample and found that
the 
galaxy luminosity obtained from their multicomponent and two-component 
decompositions are similar (Hon et al.\ 2022, PhD thesis work in preparation
for publication).

We applied the following ($g-i$ colour)-dependent stellar mass-to-light ratio prescription from
\citet[][based upon a Chabrier IMF]{Bell:2003} 
to the (Galactic extinction\footnote{Galactic extinction is obtained from
  \citet{Schlafly:2011}, available at the NASA/IPAC Extragalactic Database.}
and $K$-corrected\footnote{We obtained $K$-corrections for the SDSS-$i$ sample using a
  $K$-correction script \citep{Chilingarian:Melchior:2010,
    Chilingarian:Zolotukhin:2012} based upon the $g-i$ colour.  These are
  however tiny given that the SDSS-$i$ sample resides within 100 Mpc.}) galaxy luminosities
of the SDSS-$i$ sample to obtain the galaxy stellar mass.
\begin{equation}\label{g_i_ML}
(\log M/L)_{\rm i} = -0.152+0.518*(g-i)-0.093.
\end{equation}
As was also done in \citetalias[][]{Shankar:2016}, and thus with the SDSS-DR7 sample above, 
the subtraction of 0.093 dex --- following 
\citet[][their Equation~6]{Bernardi:2010} 
\citep[see also][]{Taylor:2011, Mitchell:2013} --- brings the diet-Salpeter IMF used in
\citet{Bell:2003} in line with the Chabrier IMF.  

The velocity dispersions for
the SDSS-$i$ sample are taken from the \textsc{HyperLeda} database.

\begin{figure}
\begin{center}
\includegraphics[clip=true,trim=10mm 08mm 20mm 20mm,width=\columnwidth]{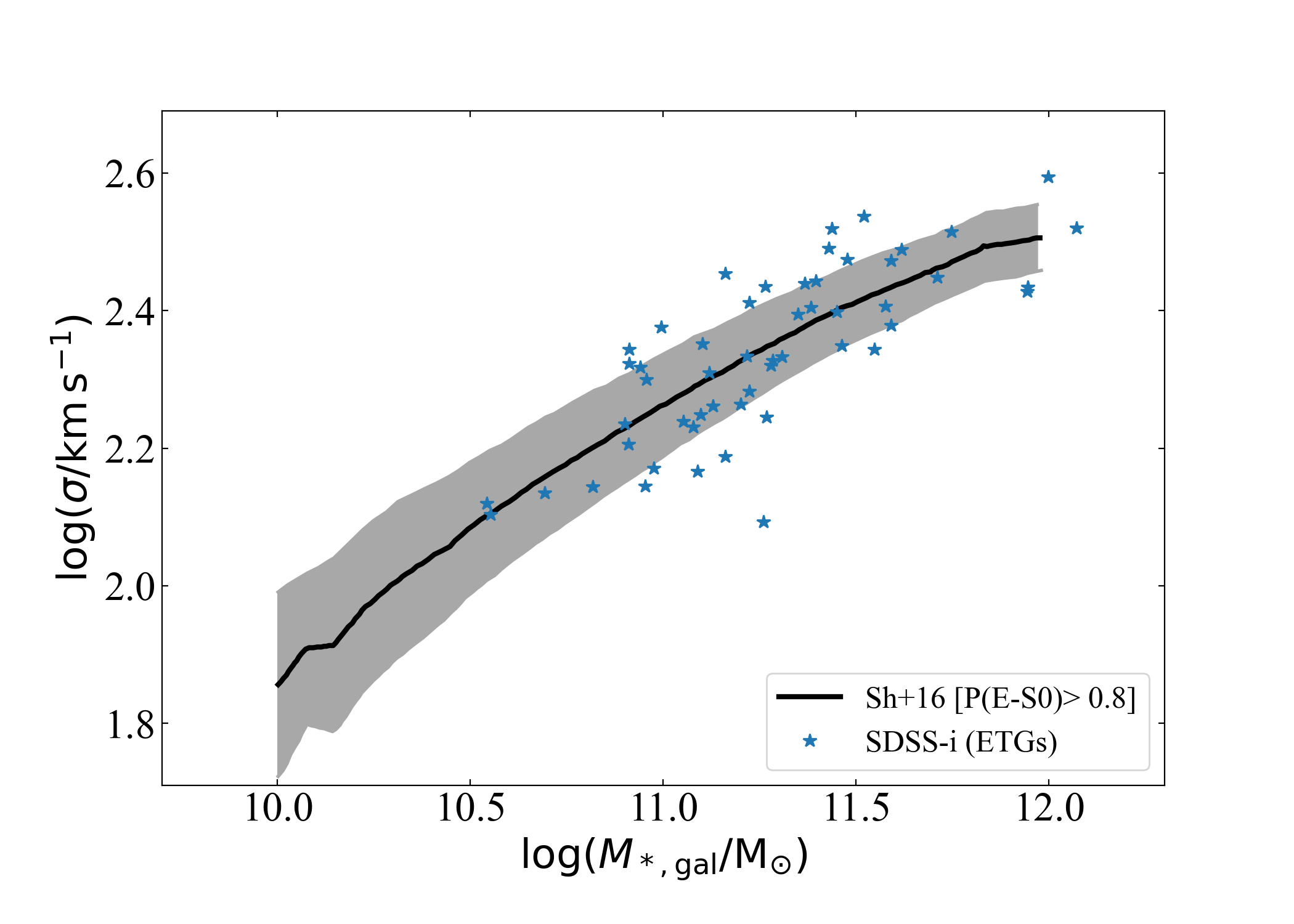} 
\caption{Same as Figure~\ref{Sigma_Mgal_sh} but now plotting the SDSS-$i$ sample over
  the $\sigma_{\rm HL}$--$M_{\rm *,gal}$ curve for the SDSS-DR7 data from
  \citetalias{Shankar:2016}.  See Section~\ref{SDSS_i}. }
\label{Sigma_Mgal_comp_SDSS}
\end{center}
\end{figure}

Figure~\ref{Sigma_Mgal_comp_SDSS} shows the 53 ETGs from this SDSS-$i$ sample
plotted along with the curve for the SDSS-DR7 ETGs from \citetalias{Shankar:2016}.
While in Figure~\ref{Sigma_Mgal_sh}, we saw the ETGs from the BHS offset from the
SDSS-DR7 distribution, here we see that the SDSS-$i$ (ETGs) are uniformly
distributed about the $\sigma_{\rm HL}$--$M_{\rm *,gal}$ curve from
\citetalias{Shankar:2016}. Given this agreement between the SDSS-DR7 and
SDSS-$i$ samples, we can use the SDSS-$i$ ETG sample from \citet{Hon:2022} 
to investigate the offset between the SMBH sample and the SDSS-DR7
sample in the $\sigma_{\rm HL}$--$M_{\rm *,gal}$ diagram.  To do so, we simply
require a calibrator sample.

\subsection{$\rm S^4G$ sample}
\label{S4G}

We used the S$^4$G catalogue of local galaxies \citep{Sheth:2010} as our
reference sample for galaxy stellar mass comparison.  The $\rm S^4G$ is a
catalogue of nearby (distance $< 40 \rm \, Mpc$) galaxies imaged by the
Infra-Red Array Camera \citep[IRAC:][]{Fazio:Hora:2004} onboard
the Spitzer Space Telescope.  The S$^4$G provided galaxy images in $3.6 \, \mu$m and
$4.5 \,\mu$m bands with outer galaxy isophotes traced out to $\rm 1 \, M_\odot
/ pc^{2}$.  The two IRAC band images are thought to provide the best estimates of
the stellar mass of a galaxy.  This is because the galaxy images/light
profiles in these bands are unaffected by young star bias (which is prominent
in optical images) or by dust obscuration \citep[although LTGs can have glowing
 warm dust:][]{Querejeta:2015}.  The (3.6-4.5 $\mu$m)
colour is almost constant with radius and independent of the stellar
population age and mass function, suggesting a stable stellar
mass-to-light ratio in these bands \citep{Jun:Im:2008, Meidt:2014}.

We use two samples of 37 and 43 ETGs from the $\rm S^4G$ catalogue
comprising the galaxies in common with the BHS (of 73 ETGs) and the 
SDSS-$i$ sample (of 53 ETGs), respectively.  The galaxy stellar mass, galaxy magnitude, and
distance information for these galaxies are taken from the publicly available
$\rm S^4G$
catalogue.\footnote{\url{https://cdsarc.unistra.fr/viz-bin/nph-Cat/html?J/PASP/122/1397/s4g.dat.gz}}
The galaxy stellar mass for the $\rm S^4G$ sample is derived using the
3.6~$\mu$m magnitudes, obtained from four-component (bulge, disk, bar, and
nuclear component, where required) fits to the galaxy light, and a ($3.6 - 4.5
\, \mu$m) colour-dependent mass-to-light ratio with a mean (median and mode)
value of $\sim 0.6$, consistent with the BHS (Section \ref{BHS}), based upon
the Chabrier IMF.  The central velocity dispersions for the $\rm S^4G$ samples
used here are also taken from the \textsc{HyperLeda} database.


\section{Analysis}
\label{analysis}



\subsection{Galaxies with and without dynamically measured SMBHs in the $\sigma$--$M_{\rm *,gal}$ diagram}\label{Sec_analysis}

\begin{figure}
\begin{center}
\includegraphics[clip=true,trim= 10mm 08mm 20mm 20mm,width=\columnwidth]{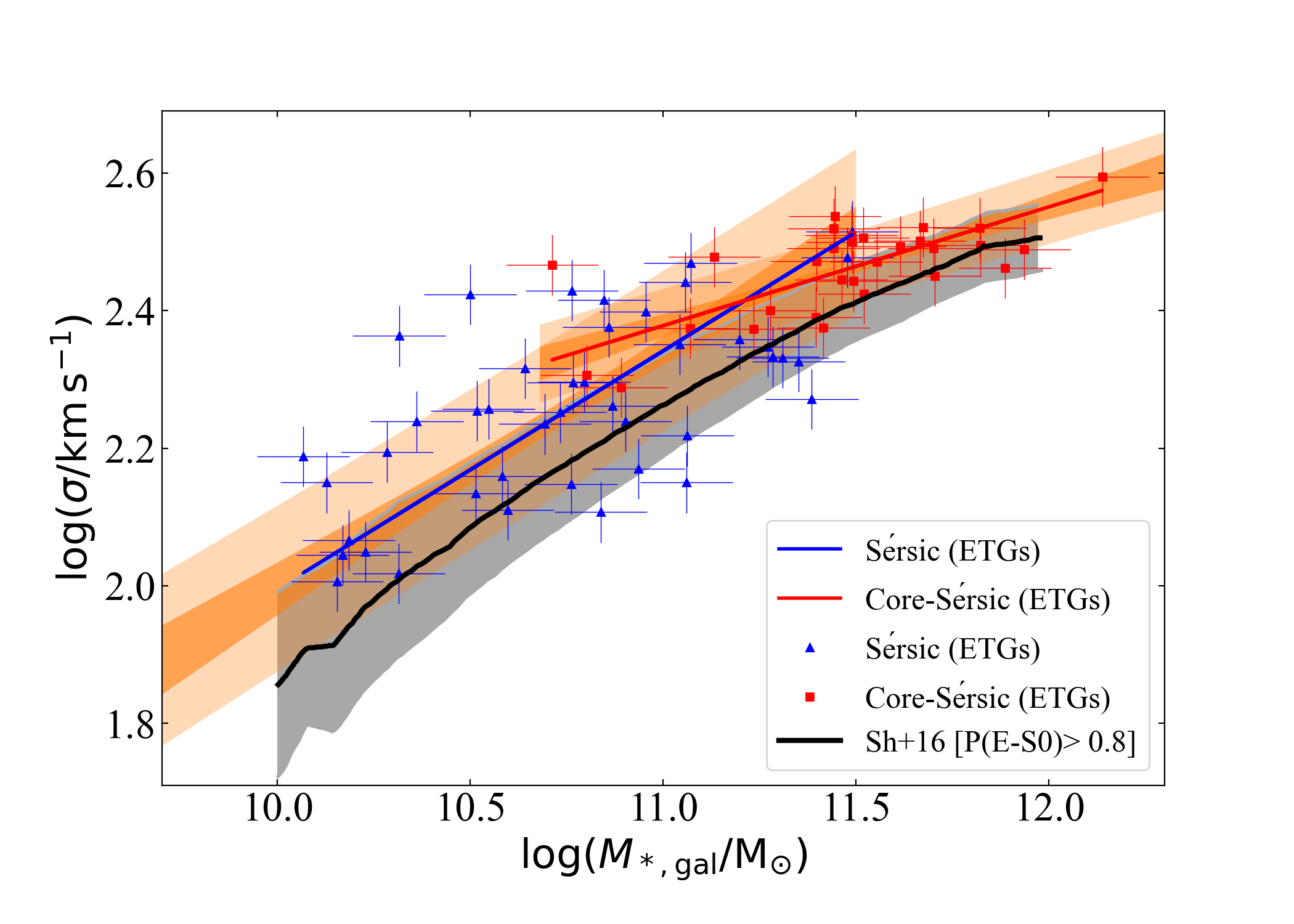} 
\caption{The $\sigma_{\rm HL}$--$M_{\rm *,gal}$ diagram \citep[adapted
    from][]{Sahu:2019:II} showing the two power-law relations defined by
  S\'ersic ETGs (blue triangles) and core-S\'ersic (red squares) galaxies in
  the BHS.  The best-fit relations and interpretation of the orange-shaded
  regions are provided in the text.  The black curve for the SDSS-DR7 data
  from \citetalias{Shankar:2016} is plotted here for comparison with the
  updated BHS.  It can be seen that the $\sigma_{\rm HL}$--$M_{\rm *,gal}$
  curve for the SDSS-DR7 data has some overlap and reduced offset (compare
  Figure~\ref{Sigma_Mgal_sh}) with the broken $\sigma_{\rm HL}$--$M_{\rm
    *,gal}$ relations defined by the updated BHS (see
  Section~\ref{Sec_analysis}.}
\label{Sigma_Mgal_comp1}
\end{center}
\end{figure}

Figure~\ref{Sigma_Mgal_comp1} shows the latest (bent) $\sigma_{\rm
  HL}$--$M_{\rm *,gal}$ relations obtained using the ETGs in the BHS
(Section~\ref{BHS}), where S\'ersic\footnote{S\'ersic galaxies likely grow through
  gas abundant mergers/accretion, and their bulge/spheroid light profile is
  aptly described using a \citet{Sersic:1968} function.} ETGs (blue triangles)
and core-S\'ersic\footnote{Core-S\'ersic galaxies, unlike the S\'ersic
  galaxies, have a depleted central bulge/spheroid light profile which is
  described using the core-S\'ersic function \citep{Graham:2003:CS}. The deficit
  of stellar light is caused during dry (gas-poor)
  merger-driven evolution, where the inspiral of the central SMBHs of the
  merging galaxies scours out stars from the centre by 
  orbital angular momenta transfer \citep{Begelman:1980}.}  galaxies (red
squares), all of which are ETGs, define distinct relations.  This $\sigma_{\rm
  HL}$--$M_{\rm *,gal}$ diagram is adapted from \citet[][their
  Figure~15]{Sahu:2019:II} with the best-fits obtained using the symmetric
Bivariate Correlated Errors and Intrinsic Scatter (\textsc{bces}) linear
regression \citep{Akritas:Bershady:1996}.  We continue using the symmetric
\textsc{bces} (aka, \textsc{bces(bisector)}) regression for all the fits shown
in this paper.  We checked each fit by performing a symmetrical analysis of
the data using two additional routines: {\sc linmix} \citep{Kelly:2007} and
{\sc mfitexy} \citep{Tremaine:ngc4742:2002}.  While these are asymmetrical
regression routines which assume that one variable is dependent on the other,
swapping the order of the two variables input into these codes provides
a second relation that can be combined with the first to yield a symmetrical
bisector regression.  In doing so, we obtained consistent results from all three
codes.

The relations defined by the S\'ersic ETGs (blue line)  and core-S\'ersic ETGs
(red line) galaxies are, respectively:
\begin{equation}\label{Ser}
\log \sigma_{\rm HL} = (0.40\pm 0.05) \log (M_{\rm *,gal}/10^{11}\,{\rm M}_\odot) + (2.35\pm 0.02),
\end{equation}
with a total root-mean-square (rms) scatter in the vertical direction ($\Delta_{\rm rms|\sigma_{\rm HL}}$) of 0.16 dex, and 
\begin{equation}\label{core-Ser}
\log \sigma_{\rm HL} = (0.17\pm 0.03) \log (M_{\rm *,gal}/10^{11}\,{\rm M}_\odot) + (2.38\pm 0.02),
\end{equation}
with $\Delta_{\rm rms|\sigma_{\rm HL}}$=0.05 dex. Here, we used a 0.12 dex
uncertainty on $\log M_{\rm *,gal}$ and a 10 per cent uncertainty on
$\sigma_{\rm HL}$ \citep[following][]{Sahu:2019:I, Sahu:2019:II}.  The dark
orange shaded regions around the blue and red lines shown in
Figure~\ref{Sigma_Mgal_comp1} mark the 68 per cent certainty on the slopes and
intercepts, and the light orange shaded regions enclose 68 per cent of the data
distributed about these lines.  The meaning of the dark and light-shaded
regions around the best-fit line remains the same throughout the paper.

Although the bend or curvature in the $\sigma$--$M_{\rm *,gal}$ (or $L_{\rm
  gal}$--$\sigma$) diagram has been known for a long time
\citep[e.g.,][]{Davies:1983, Farouki:1983, deRijcke:2005,
  Matkovic:Guzman:2005, Lauer:Faber:2007, Graham:Soria:2018}, the size of the
SMBH sample used for the comparison in \citetalias{Shankar:2016} was not large
enough to see this bend within the SMBH sample.  We can now see that the bent
$\sigma$--$M_{\rm *,gal}$ relation defined by the BHS is almost parallel to
the curvature defined by the larger SDSS-DR7 sample.  One can fit a curve to
describe the distribution of the BHS in the $\sigma_{\rm HL}$--$M_{\rm *,gal}$
diagram (Figure~\ref{Sigma_Mgal_comp1}); however, we will continue using two
power-law fits for BHS as it does not affect the outcome of this
investigation.

Compared to Figure~\ref{Sigma_Mgal_sh}, Figure~\ref{Sigma_Mgal_comp1} (which
uses the current expanded BHS) shows a mildly reduced offset between the BHS
and the SDSS-DR7 sample, i.e., galaxies without a direct SMBH mass
measurement.  The $\sigma_{\rm HL}$--$M_{\rm *,gal}$ (black) curve defined by
the SDSS-DR7 sample lies within the 68 per cent ($\pm 1 \sigma$) scatter bound
(i.e., the light orange shaded region) of the BHS. The $\pm 1 \sigma$ scatter
(grey) region of the SDSS-DR7 curve also partially overlaps with the 
$\pm 1 \sigma$ uncertainty (dark orange shaded) region about the best-fit
relations for the BHS.  However, there is still a noticeable offset between
the best-fit relations for the BHS and the mean $\sigma_{\rm HL}$--$M_{\rm
  *,gal}$ curve defined by the SDSS-DR7 sample.

In the following subsections, we explore if this offset is because of a
discrepancy in the stellar mass calculation, primarily due to the stellar
mass-to-light ratio used to convert galaxy luminosity into galaxy stellar
mass.  We investigate the offset in stellar mass by comparing the BHS and
SDSS-$i$ samples (and thus the BHS and SDSS-DR7 samples, given the agreement
between the SDSS-$i$ and SDSS-DR7 samples in
Figure~\ref{Sigma_Mgal_comp_SDSS}) using the $\rm S^4G$ sample as a reference.

\subsection{Comparison between the SMBH sample and the $\rm S^4G$ sample}
\label{sec_BHS_S4G}

\begin{figure}
\begin{center}
\includegraphics[clip=true,trim= 10mm 08mm 20mm 20mm, width= \columnwidth]{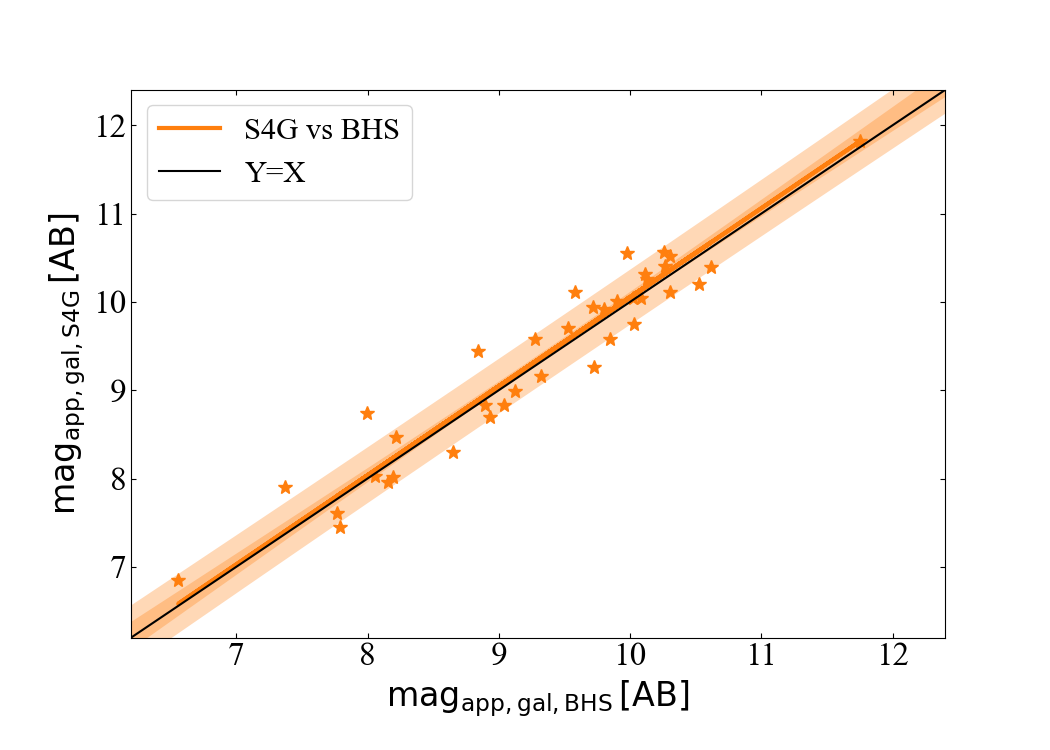}
\caption{A comparison of galaxy apparent magnitude (Spitzer 3.6~$\mu$m)
  between the BHS (horizontal-axis) and the $\rm S^4G$  sample (vertical-axis)
  using 37 galaxies common to both samples.  See Section \ref{sec_BHS_S4G}.}
\label{S4G_vs_Sahu}
\end{center}
\end{figure}


Figure~\ref{S4G_vs_Sahu} compares the BHS and the reference $\rm S^4G$ sample
using 37 galaxies common to the two samples. As both samples have $3.6\,
\mu$m-band imaging, we can directly compare the galaxy's apparent
magnitudes, shown in the top panel of Figure~\ref{S4G_vs_Sahu}.  The $y=x$
(black) line and the orange \textsc{bces(bisector)} line fit\footnote{For
  an appropriate comparison of galaxy properties (mass, magnitude) from
  different samples, the \textsc{bces(bisector)} regression is preferred here
  because this symmetric regression offers equal treatment of both variables.}
to the distribution in Figure~\ref{S4G_vs_Sahu} are provided to quickly assess
the one-to-one correspondence between the measured galaxy magnitudes, and thus
stellar masses, of galaxies in common with the two samples.



\subsection{Comparison between SDSS-$i$ sample and the $\rm S^4G$ sample}
\label{sec_SDSSi_S4G}

\begin{figure}
\begin{center}
\includegraphics[width=\columnwidth]{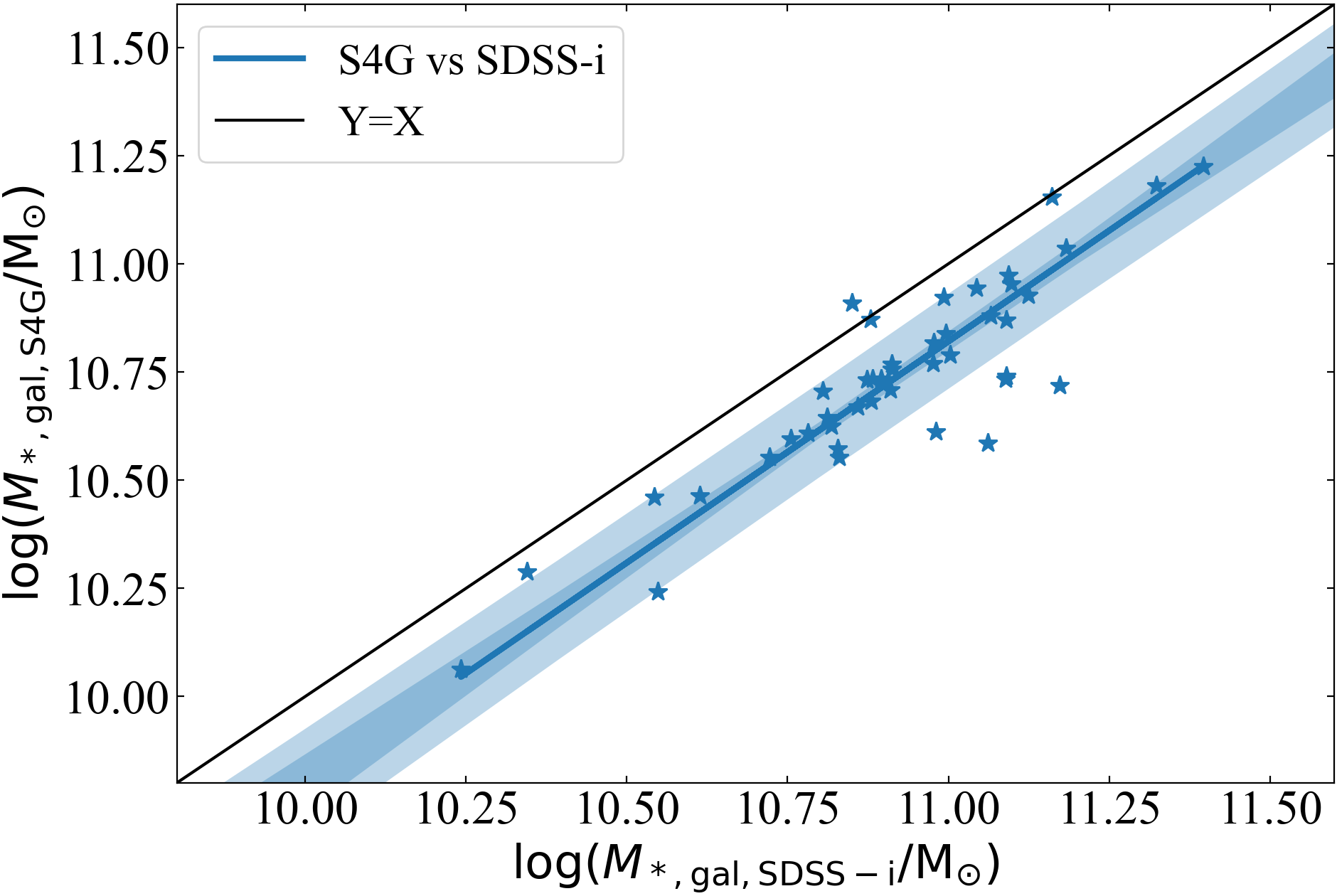}
\caption{Comparing the total galaxy stellar mass from $\rm S^4G$ and SDSS-$i$
  samples for 43 galaxies in common, using the same galaxy distances (from
  SDSS-$i$) for both.  See Section~\ref{sec_SDSSi_S4G}. }
\label{Mod_S4G_vs_Hon}
\end{center}
\end{figure}

Figure~\ref{Mod_S4G_vs_Hon} compares the galaxy stellar masses between the
$\rm S^4G$ and SDSS-$i$ samples using 43 galaxies in common to both samples.  The X-axis shows
the galaxy stellar mass based upon the SDSS $i$-band galaxy magnitudes from
\citet{Hon:2022}, and the ($g-i$ colour)-based mass-to-light ratio
(Equation~\ref{g_i_ML}), 
and the Y-axis shows the stellar mass of the same
galaxies based upon the $\rm S^4G$ 3.6~$\mu$m galaxy magnitudes and
$M/L_{3.6}=0.6$.
Given the agreement in Figure~\ref{S4G_vs_Sahu}, this will yield stellar
masses for the S$^4$G sample that are consistent with the BHS.  
%
%
Given the agreement in Figure~\ref{Sigma_Mgal_comp_SDSS}, it will also yield SDSS-$i$
stellar masses that are consistent with SDSS-DR7 from \citetalias{Shankar:2016}. 
Here, for both $M_{\rm
  \tiny{*,gal,SDSS-i}}$ and $M_{\rm \tiny{*,gal,S^4G}}$, we adopted the
galaxy distances from \citet{Hon:2022}. 

Figure~\ref{Mod_S4G_vs_Hon} shows that the SDSS-$i$ sample has a higher
galaxy stellar mass than the $\rm S^4G$ sample.  In
Figure~\ref{Sigma_Mgal_comp1}, we saw that the SDSS-DR7 sample masses used in
\citetalias{Shankar:2016}, which agree with the SDSS-$i$ sample
(Figure~\ref{Sigma_Mgal_comp_SDSS}), are offset from the BHS.  At the same
time, the galaxy stellar masses from the $\rm S^4G$ sample and the BHS
(Figure~\ref{S4G_vs_Sahu}) agree well; thus, this offset between the stellar
masses from SDSS-$i$ and $\rm S^4G$ was expected.

Given that we are comparing the entire galaxy stellar mass rather than the mass of
a galaxy component, and both \citet[][for the SDSS-$i$ sample]{Hon:2022} and
\citet[][for the $\rm S^4G$ sample]{Sheth:2010} performed multicomponent
decomposition, rather than single S\'ersic fits, a discrepancy in galaxy
magnitude should not be the dominant reason behind this offset.  This suggests
that the significant offset between the stellar masses for the SDSS-$i$ and
$\rm S^4G$ masses is because of an inconsistency in the stellar mass-to-light
ratio prescriptions used.

To quantify the offset between the SDSS-$i$ and $\rm S^4G$ stellar masses, we
perform a symmetric regression providing the best-fit (blue) line shown in
Figure~\ref{Mod_S4G_vs_Hon}. This line can be expressed as
\begin{eqnarray}\label{S4G_SDSSi_mass}
\log \left( \frac{M_{\rm \tiny{*,gal,S^4G}}}{{\rm M}_\odot}
\right) & = & (1.02\pm0.05) \log \left( \frac{M_{\rm \tiny{*,gal,SDSS-i}}}{10^{11}\,
  {\rm M}_\odot} \right) \nonumber \\
 & & + (10.82\pm 0.02).
\end{eqnarray}
The mean horizontal offset ($\Delta_{\rm S^4G -SDSS-i} \equiv \log (M_{\rm
  *,gal,S^4G}) - \log (M_{\rm *,gal,SDSS-i})$) of the blue line in
Figure~\ref{Mod_S4G_vs_Hon} as a function of the (SDSS-$i$)-based mass is 
\begin{equation}\label{cor_S4G_SDSSi_mass}
\Delta_{\rm S^4G -SDSS-i} = 0.02 \log \left( \frac{M_{\rm *,gal,
    SDSS-i}}{10^{11} {\rm M}_\odot } \right) - 0.18. 
\end{equation}

\subsection{Correcting the $\sigma$--$M_{\rm *,gal}$ curve}
\label{correction}

Discrepancies in stellar mass-to-light ratio have been frequently discussed in
the literature \citep[e.g.,][]{Kannappan:Gawiser:2007, Taylor:2011,
  Zhang:Puzia:2017, Graham:Soria:Davis:2018, Schombert:McGaugh:Lelli:2019,
  Lower:Narayanan:2020}, including \citet{Davis:2018:a} and
\citet{Sahu:2019:I}.  Using an inconsistent stellar mass in the observed
$M_{\rm bh}$--$M_{\rm *,gal}$ relation will predict an erroneous SMBH mass.
Thus, \citet{Davis:2018:a} introduced a mass correction coefficient
($\upsilon$), defined as a ratio of the stellar mass-to-light ratio used to
construct the SMBH scaling relations and to measure the mass of a new
target of interest.  Essentially, this correction coefficient brings stellar
masses calculated using another mass-to-light ratio prescription into 
consistency with the scaling relations.

\citet[][in their Section 3.4]{Sahu:2019:I} generalised this mass
correction coefficient to accommodate stellar masses based upon luminosity measured in
different bands and as a function of the stellar mass.  The $\Delta_{\rm S^4G
  -SDSS-i}$ derived here can be thought of as the correction coefficient $\log
\upsilon_{*,\rm SDSS-i}$, which should be added to the SDSS-$i$ stellar masses
(SDSS-$i$ magnitude + \citet{Bell:2003} mass-to-light ratio $+$
\citet{Chabrier:2003} IMF) to bring them into consistency with the $\rm S^4G$
stellar masses.  This approach of using $\upsilon$ in the SMBH scaling
relations involving stellar mass to resolve the stellar mass discrepancy is
not yet widespread, given the challenge of estimating $\upsilon$ for all
possible filters plus mass-to-light ratio combinations.  However, it is
crucial.  

From the above investigations, we found the following.
\begin{itemize}
\item Our SDSS-$i$ and the SDSS-DR7 masses agree in the
$\sigma$--$M_{\rm *,gal}$ diagram (Figure~\ref{Sigma_Mgal_comp_SDSS}). 
\item Both  SDSS-$i$ and SDSS-DR7 masses are offset 
towards higher $M_{\rm *,gal}$ or lower $\sigma$ 
from the BHS in the $\sigma$--$M_{\rm *,gal}$ diagram (Figure~\ref{Sigma_Mgal_comp1}).
\item There is a good agreement of BHS galaxy magnitudes, and thus 
stellar masses, with the reference $\rm S^4G$ sample (Figure~\ref{S4G_vs_Sahu}). 
\item There is a  systematic difference between SDSS-$i$ (and thus SDSS-DR7) 
and the  $\rm S^4G$ galaxy stellar masses, where the SDSS-$i$ 
stellar masses are systematically higher than the reference $\rm S^4G$ sample
(Figure~\ref{Mod_S4G_vs_Hon}). This offset is quantified by Equation
\ref{cor_S4G_SDSSi_mass}. 
\item This offset in the galaxy stellar mass likely originates from the inconsistency in the adopted (stellar mass)-to-light ratio prescriptions.
\end{itemize}

The offset of the SDSS-$i$ masses (which agrees with the SDSS-DR7 masses) from
the reference $\rm S^4G$ sample (which agrees with the BHS), is also a measure
of the offset between the SDSS-DR7 and BHS masses.  Thus, the use of Equation
\ref{cor_S4G_SDSSi_mass} enables us to correct for the differing/inconsistent
derivations of the stellar mass, which may explain the offset seen in the
$\sigma$--$M_{\rm *,gal}$ diagram (Figure~\ref{Sigma_Mgal_comp1}) between the
relations defined by the BHS and \citetalias{Shankar:2016} curve formed by the
SDSS-DR7 ETG sample.

\begin{figure}
\begin{center}
\includegraphics[width= \columnwidth]{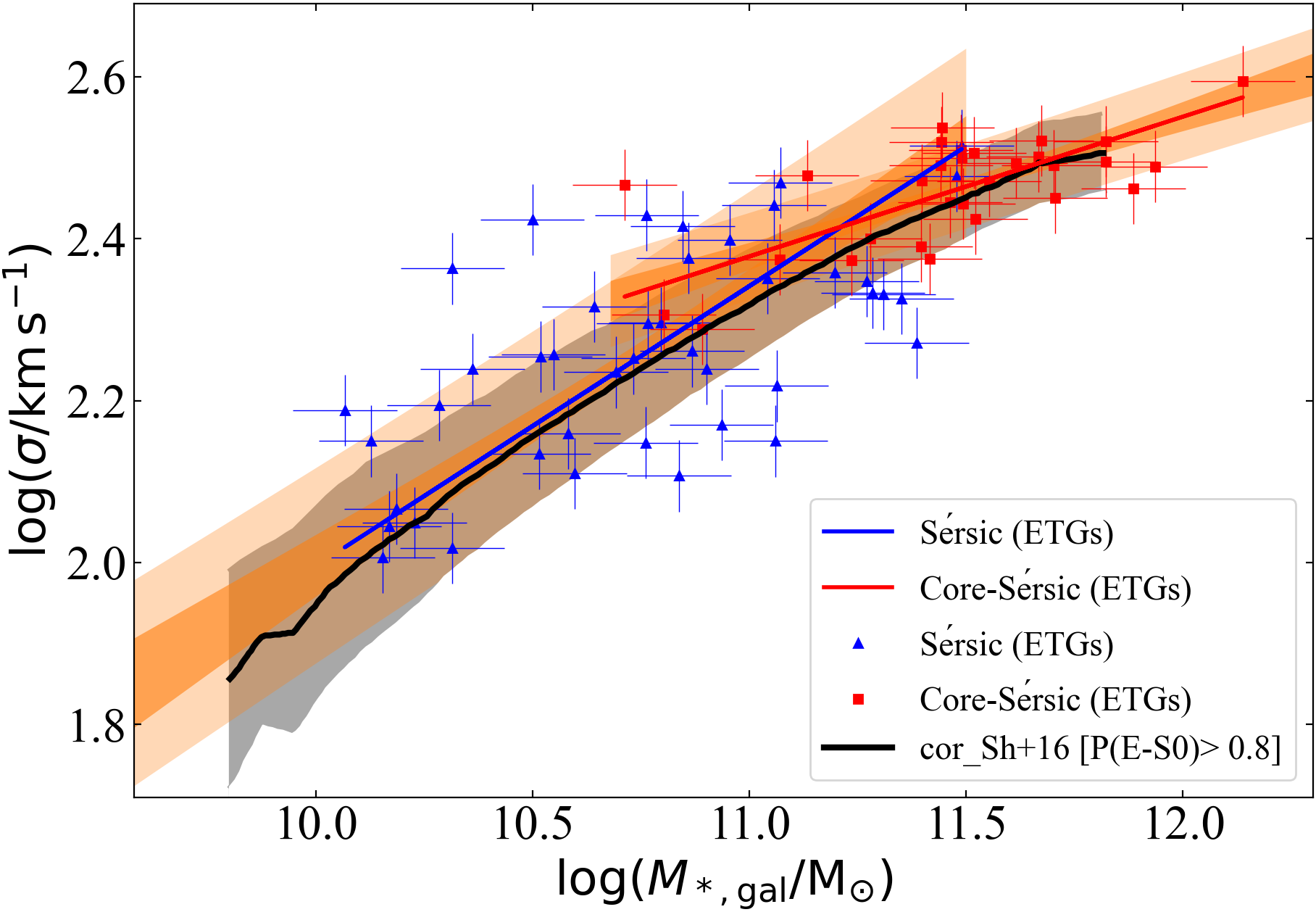}
\caption{Similar to Figure~\ref{Sigma_Mgal_comp1}, but now showing the mass discrepancy corrected $\sigma_{\rm HL}$--$M_{\rm *,gal}$ curve form by ETGs in the SDSS-DR7 ETG sample.}
\label{Mass_Sigma_corrected}
\end{center}
\end{figure}

Figure~\ref{Mass_Sigma_corrected} presents the $\sigma_{\rm HL}$--$M_{\rm
  *,gal}$ diagram with the systematic stellar mass correction (i.e., the
addition of the mean logarithmic offset $\Delta_{\rm S^4G -SDSS-i} $, Equation
\ref{cor_S4G_SDSSi_mass}) applied to the SDSS-DR7 galaxy masses.  Evidently,
the corrected black curve defined by the SDSS-DR7 ETG sample now matches very
well with the bent $\sigma_{\rm HL}$--$M_{\rm *,gal}$ relations defined by
S\'ersic and core-S\'ersic ETGs in the BHS.  The correction provided by
Equation~\ref{cor_S4G_SDSSi_mass} translates into a 0.1 dex shift in
$\log(\sigma)$ in the $\sigma$--$M_{\rm *,gal}$ diagram. For a slope of
$\sim$5 in the $M_{\rm bh}$--$\sigma$ diagram, this maps to a shift 
of $\sim$0.5 dex in the SMBH mass.

\section{Implications for the Sh$+$16 ``intrinsic models''}
\label{Implication for BH}


\citetalias{Shankar:2016} believed that the offset initially seen in the
$\sigma_{\rm HL}$--$M_{\rm *,gal}$ diagram is natural and occurs because, for
the same galaxy stellar mass, the BHS has a higher $\sigma$ relative to the
broader SDSS-DR7 sample.  That is, they thought that the SMBH scaling
relations are selection biased as they are based on a sample of local galaxies
with massive SMBHs having higher $\sigma$ than the population at large.  To
obtain what they thought was the unbiased SMBH scaling relation (that they
refer to as the ``intrinsic" relation) using their Monte Carlo simulation,
\citetalias{Shankar:2016} used SMBH predictor models, 
e.g., $\log M_{\rm bh}/ {\rm M}_{\odot} = \gamma + \beta \log (\sigma/ \rm
200\, km\, s^{-1}) + \alpha \log M_{\rm star}$, 
based upon the available SMBH scaling relations at the time.  Additionally,
what is important to note is that they reduced the normalisation (intercept,
$\gamma$) of their SMBH predictor models by 0.4-0.6 dex (see their Section
4.1) due to the offset seen in Figure~\ref{Sigma_Mgal_sh} but explained in
Figure~\ref{Mod_S4G_vs_Hon}.  This played an important role in producing their
``intrinsic" SMBH scaling relation, which had lower normalisation than the
(original) directly observed SMBH scaling relations, i.e., the relations based
upon a galaxy sample with SMBH masses directly measured using stellar or gas
dynamical modelling or megamaser kinematics.

\citetalias{Shankar:2016} did not specifically mention the exact reference for
their SMBH predictor models.  Based on the description and references in
their Section 4.1, the coefficients ($\alpha, \beta, \gamma$) of their SMBH
predictor models are taken from the parameters of concurrent SMBH scaling
relations \citep[e.g.,][]{Kormendy:Ho:2013, McConnell:Ma:2013}.  For example,
their Model-II ($\log M_{\rm bh} = 7.75 + 2.5\log (\sigma/200\, {\rm km}\,
{\rm s}^{-1}) + 0.5\log(M_{\rm STAR}/10^{11}\,{\rm M}_\odot$) is essentially a
linear combination of the $M_{\rm bh} \propto \sigma^{\sim 5}$ and $M_{\rm bh}
\propto M_{\rm *,bulge}^{\sim 1}$ relations from \citet{McConnell:Ma:2013} and
\citet{Kormendy:Ho:2013}, respectively, accompanied by a normalisation
($\gamma$) reduction of $\sim 0.5$ dex.  This linear combination is also very
similar to the $M_{\rm bh} \propto \sigma^{2.18} M_{\rm *}^{0.54}$ plane
provided by \citet{Hopkins:Hernquist:2007}, another reference
mentioned in \citetalias{Shankar:2016}.  In their Model-I, they doubled the
coefficient of the $\log \sigma$ term to obtain a stronger
intrinsic dependence on $\sigma$.


The reduction in the normalisation, $\gamma$, of their SMBH predictor models
by 0.4-0.6 dex was to account for an average offset of $\sim 0.1$ dex in $\log
\sigma_{\rm HL}$ (that they noticed in the $\sigma$--$M_{\rm *,gal}$ diagram),
coupled with a slope of $\sim$5 for the $M_{\rm bh}$--$\sigma_{\rm HL}$
relation.  This is because  
$\Delta \log\sigma \sim 0.1$ dex corresponds to $\Delta \log M_{\rm bh} \sim
0.5$ dex based upon $M_{\rm bh} \propto \sigma_{\rm HL}^5$ (that is,
$\Delta\log M_{\rm bh} \propto 5\Delta \log\sigma $).  This artificial
normalisation reduction in the \citetalias{Shankar:2016} SMBH predictor models
fully explains the offset seen (in $M_{\rm bh}$-direction) when comparing
their predicted $M_{\rm bh}$--$\sigma_{\rm HL}$ relation with the relation
obtained using the directly-dynamically measured SMBH sample in
\citet{Savorgnan:Graham:2015}, \citet{Nicola:2019}, and \citet{Sahu:2019:II}.
Moreover, this normalisation reduction explains a significant part of the
offset seen \citep[e.g.,][]{Shankar:Allevato:2020, Shankar:Weinberg:2020,
  Allevato:Shankar:2021} when comparing their predicted $M_{\rm bh}$--$M_{\rm
  *,gal}$ (or $M_{\rm STAR}$ ) curve with the directly observed $M_{\rm
  bh}$--$M_{\rm *,gal}$ relations based upon the dynamically measured SMBH
sample \citep{Kormendy:Ho:2013, Savorgnan:2016:Slopes, Davis:2018:b,
  Sahu:2019:I}. 

A simple increment of 0.4-0.6 dex (reversing what \citetalias{Shankar:2016}
subtracted) in the normalisation of their Monte-Carlo generated $M_{\rm
  bh}$--$\sigma_{\rm HL}$ relation \citepalias[Equation 7 in ][based upon
  their Model-I with stronger intrinsic dependence on $\sigma$]{Shankar:2016},
places it directly on top of the observed $M_{\rm bh}$--$\sigma^{5.7}$
relations in, for example, \citet{McConnell:Ma:2013},
\citet{Savorgnan:Graham:2015}, 
and the latest morphology-dependent $M_{\rm bh}$--$\sigma_{\rm
  HL}$ relations presented in \citet{Sahu:2019:II}, where normal S\'ersic
galaxies yield $M_{\rm bh}$--$\sigma^{5.75\pm 0.34}$ and the 
core-S\'ersic galaxies yield $M_{\rm bh}$--$\sigma^{8.64\pm
  1.10}$ \citep[as also seen in][]{Bogdan:2018, Dullo:GildePaz:2020}.

In the $M_{\rm bh}$--$M_{\rm *,gal}$ diagram, simply removing the
normalisation adjustment of 0.4-0.6 dex from the \citetalias{Shankar:2016}
``intrinsic" relation will bring it closer to the current, directly observed
$M_{\rm bh}$--$M_{\rm *,gal}$ relations based on the BHS
\citep{Sahu:2019:I}, but not exactly match.  This is because, first,
the \citetalias{Shankar:2016} 
$M_{\rm bh}$ was predicted using models based upon the previously current,
i.e., contemporaneous, 
$M_{\rm bh}$--$\sigma$ relation plus a linear $M_{\rm bh}$--$M_{\rm *,bulge}$
relation (with an artificially reduced normalisation). However, the $M_{\rm
  bh}$--$M_{\rm *,bulge}$ relation was incorrectly used as an $M_{\rm
  bh}$--(total galaxy) stellar mass ($M_{\rm STAR}$, or $M_{\rm *,gal}$)
relation for ETGs, which are not only comprised of purely spheroidal
elliptical galaxies but also include ellicular/lenticular (ES/S0-type) 
galaxies with discs. The second reason is the inconsistency in the calculated galaxy
stellar mass (as revealed here between the BHS and SDSS-DR7 sample), which was used in
their models to predict the SMBH mass and to generate their
``intrinsic'' SMBH relation.


Although a moot exercise now, an alternative approach would have been to
acquire velocity dispersions for the ETGs in the S$^4$G. 
This approach would have enabled the construction of the $\sigma$--$M_{\rm *,gal}$ relation
using the same infrared luminosities as done with the sample of galaxies
having directly-measured black hole masses, bypassing the (as it turned out,
biased) masses in the SDSS.

In passing, we note that one ought to use the morphology-dependent $M_{\rm
  bh}$--$M_{\rm *,gal}$ relations  --- see \citet{Sahu:2019:I} and
\citet{Graham:Sahu:2022} --- to construct the $M_{\rm bh}$ predictor model when
involving galaxy stellar mass ($M_{\rm *,gal}$) as one of the
independent/input parameters.
While the conclusion in \citetalias{Shankar:2016} is not correct, and the
sample of galaxies with directly measured SMBH masses is not biased, the
misperceived offset in the $\sigma$--$M_{\rm *,gal}$ diagram has seen some
partial success in subsequent works. This is because of the way the
``intrinsic'' model from
\citetalias{Shankar:2016} sometimes effectively readjusts the observed SMBH scaling
relation for consistency with samples whose stellar masses were derived
inconsistently with the $M_{\rm bh}$--$M_{\rm *}$ scaling relations.  However,
in application, it is not always as straight forward as this, and past work
based on the biased ``intrinsic'' $M_{\rm bh}$--$\sigma$ relation should be
revisited using the observed SMBH scaling relations.



\section{Summary}
\label{conclusions}

We have investigated the offset \citetalias{Shankar:2016} observed in the
$\sigma_{\rm HL}$--$M_{\rm *,gal}$ diagram between galaxies with dynamical
SMBH mass measurements and a sample of galaxies (SDSS-DR7,
Section~\ref{SDSS_DR7}) without dynamical SMBH mass measurements (see
Figure~\ref{Sigma_Mgal_sh}).  As seen in Figure~\ref{Sigma_Mgal_comp1}, while
the current sample of ETGs with dynamical SMBH mass measurements (BHS:
Section~\ref{BHS}) has more data points overlapping with the SDSS-DR7 sample
than compared to a subset from \citet{Savorgnan:2016:Slopes}, the distinct
$\sigma_{\rm HL}$--$M_{\rm *,gal}$ relations defined by the S\'ersic and
core-S\'ersic galaxies from the BHS are offset and almost parallel to the
$\sigma_{\rm HL}$--$M_{\rm *,gal}$ curve defined by the SDSS-DR7 sample of
ETGs.  At odds with the assumption in \citetalias{Shankar:2016}, our
investigation has revealed that this offset in the $\sigma_{\rm HL}$--$M_{\rm
  *,gal}$ diagram is in the horizontal direction; that is, the offset is due
to a discrepancy in the derived galaxy stellar mass rather than different the velocity
dispersions.

The $\sigma$--$M_{\rm *,gal}$ diagram shows 
no natural offset/discrepancy between the BHS and galaxies without
dynamical SMBH mass measurements.  The
SMBH scaling relations defined by galaxies with dynamically measured SMBH
masses are not biased.  The initial offset seen by \citetalias{Shankar:2016}
and, further, in Figure~\ref{Sigma_Mgal_comp1}, is artificial and caused
by an inconsistency in the derivation of the stellar mass.
This stellar-mass offset is quantified in Equation \ref{cor_S4G_SDSSi_mass},
which can be used as a mass correction equation to bring the SDSS-DR7 galaxy
masses into agreement with the $\rm S^4G$ masses and thus the BHS masses. 

We conclude that one does not have to account for a bias/offset when correctly
applying various directly observed BH--galaxy correlations. For example, to
predict $M_{\rm bh}$ in other galaxies, to calibrate the virial factor
required to convert the virial product of an active galactic nucleus (AGN) into an SMBH mass via the
reverberation mapping technique \citep{Bennert:2011, Bentz:2015}, to estimate
the SMBH mass function of the Universe \citep{McLure:Dunlop:2004,
  2007MNRAS.380L..15G, Kelly:2007}, and to estimate the detectable/expected
long gravitational-wave background (GWB) strain and event rate
\citep{Shannon:2015, Sesana2016}.  However, to correctly use the SMBH scaling
relations for the above purposes, one should account for the possible
discrepancy in the derivation of the stellar mass (or other input parameters)
and also note the galaxy morphology when using the latest morphology-dependent
SMBH scaling relations.


Our conclusion further suggests that the correction in the expected GWB
amplitude estimated in \citet{Sesana2016}, accounting for the bias in SMBH
scaling relations claimed by \citetalias{Shankar:2016}, was not required.
However, we strongly recommend revising the expected GWB models using the
latest morphology-dependent SMBH scaling relations.  These can significantly
modify the previous estimates based upon the old linear $M_{\rm bh}$--$M_{\rm
  *,bulge}$ relation and an assumed constant bulge-to-total stellar mass ratio
\citep[e.g.,][]{Shannon:2015}.  \citet[][Section 5]{Sahu:2022} provide more
details on applying the latest SMBH scaling relations for an
improved GWB strain model.  Similarly, the morphology-dependent SMBH scaling
relations can now provide improved morphology-aware virial factors,
morphology-aware SMBH mass functions, better SMBH mass estimates for galaxies
with known morphology, and improved tests for simulations hoping to form
realistic galaxies with central SMBHs.




\section*{Acknowledgements}

We thank the astrophysics group at the University of Queensland for hosting
N.S.\ for over a year during the covid-19 pandemic.
Some of this research was conducted within the Australian Research Council's
Centre of Excellence for Gravitational Wave Discovery (OzGrav), through
project number CE170100004. This project was also 
supported under the Australian Research Council's funding scheme DP17012923. 
This work has made use of the NASA/IPAC Infrared Science Archive
and the NASA/IPAC Extragalactic Database (NED), which are 
funded by NASA and operated by the California Institute of Technology. 
Funding for the SDSS was provided by the Alfred P.\ Sloan
Foundation, the Participating Institutions, the National Science Foundation,
the U.S.\ Department of Energy, the National Aeronautics and Space
Administration, the Japanese Monbukagakusho, the Max Planck Society, and the
Higher Education Funding Council for England. 
This research has made use of NASA's Astrophysics Data System Bibliographic
Services and the \textsc{HyperLeda} database (\url{http://leda.univ-lyon1.fr}).
After completing the analysis for this paper just prior to the pandemic and
extensive lockdowns in Melbourne, we discovered that
\citet{2022AJ....163..154S} also offers a self-consistent prescription for
$g-i$ and $V-$[3.6] colour-dependent stellar mass to light ratios which will
be useful for some readers.

\section*{Data Availability}
The imaging data can be found from the sources referenced in
Section~\ref{Data}, and includes 
SDSS-DR8 (\url{https://www.sdss3.org/dr8/}),
S$^4$G (\url{https://irsa.ipac.caltech.edu/data/SPITZER/S4G/}), and 
the Spitzer Heritage Archive (SHA: \url{http://sha.ipac.caltech.edu}). 
Concise data tables can also be requested from N.S.

\bibliographystyle{mnras}
\bibliography{bibliography}
\bsp	
\label{lastpage}
\end{document}